\documentclass[twocolumn]{revtex4}
\usepackage[dvipdfmx]{graphicx}
\usepackage{amsmath,amsbsy,amssymb}
\usepackage{bm}
\usepackage{mathrsfs}
\usepackage{ulem}
\usepackage{textgreek}

\usepackage{color}

\newcommand{\diff}{\mathrm{d}}
\newcommand{\imag}{\mathrm{Im}\,}
\newcommand{\real}{\mathrm{Re}\,}
\newcommand{\trace}{\mathrm{Tr}\,}
\newcommand{\imu}{\mathrm{i}}
\newcommand{\epn}{\mathrm{e}}

\newcommand{\dg}{\dagger}
\newcommand{\la}{\langle}
\newcommand{\ra}{\rangle}
\newcommand{\al}{\alpha}
\newcommand{\sg}{\sigma}
\newcommand{\gm}{\gamma}
\newcommand{\ep}{\varepsilon}

\begin{document}

\title{
Unconventional 
orbital
ordering and emergent dimensional reduction
\\
in fulleride superconductors
}

\author{Shintaro Hoshino$^1$, Philipp Werner$^2$, and Ryotaro Arita$^{3,4,5}$}

\affiliation{
$^{1}$Department of Physics, Saitama University, Shimo-Okubo, Saitama 338-8570, Japan
\\
$^{2}$Department of Physics, University of Fribourg, 1700 Fribourg, Switzerland
\\
$^{3}$Department of Applied Physics, University of Tokyo, Hongo, Tokyo 113-8656, Japan
\\
$^{4}$RIKEN Center for Emergent Matter Science (CEMS), Wako, Saitama 351-0198, Japan
\\
$^{5}$JST, ERATO, Isobe Degenerate $\pi$-Integration Project, Hongo,
Tokyo 113-0033, Japan
}

\date{\today}

\begin{abstract}
Non-local order parameters in space-time are proposed to characterize the unconventional orbital-selective conducting state in fulleride superconductors, called the Jahn-Teller metal.
In previous works, it has been argued that this state can be interpreted as a spontaneous orbital-selective Mott state, in which the electrons in two of the three $t_{1u}$ molecular orbitals are localized, while those in the third one are metallic.
Here, based on the realistic band structure for fullerides, we provide a systematic study of nonlocal order parameters and characterize the Jahn-Teller metal,
for which there exists 
no one-body local order parameter in contrast to conventional orderings.
It is shown that the Mottness, or integer filling nature for each orbital due to strong correlation effects,
is a relevant feature of 
the present orbital order.
The local orbital moment thus vanishes and 
the static distortion associated with a conventional orbital moment is absent. 
Transport characteristics are also investigated, and it is found that the dimensionality is effectively reduced from three to two at low energies, while the cubic nature is recovered at high energies.
This accounts for the high upper critical field observed in the superconducting state of the fcc fullerides inside the Jahn-Teller metal regime.
\end{abstract}

\maketitle

\section{Introduction}

Unconventional superconducting states are found in a broad range of materials with $p$, $d$, or $f$ electrons, and the exploration of the complex physical properties of these compounds is a central topic in condensed matter physics.  Fullerene-based superconductors \cite{Hebard91,Rosseinsky91,Holczer91,Tanigaki91,Fleming91},  
which exhibit a superconducting dome 
in the vicinity of a Mott insulating phase, are an interesting example  \cite{Ganin08,Takabayashi09,Capone09,Ihara10,Ganin10,Nomura16}. In $A_3$C$_{60}$, three electrons are doped onto each fullerene molecule from intercalated alkaline metals denoted by $A$ \cite{Gunnarsson97}. In the metallic compounds, the $t_{1u}$ orbitals form three half-filled narrow bands, and the electronic correlations are strong \cite{Nomura12}. Even though the symmetry of the pairing state is $s$-wave, 
the superconductivity is
different from that of conventional BCS superconductors \cite{Capone02,Capone09}. An important ingredient in the superconducting mechanism is an effectively sign-reversed Hund's coupling \cite{Fabrizio97,Capone00}. This antiferromagnetic Hund's coupling favors the low-spin ($S=1/2$) state rather than the high-spin ($S=3/2$) state, which is favored by the usual Hund's rule coupling. The low-spin state has doubly occupied orbitals, which can act as a seed for superconductivity \cite{Capone01,Capone02,Capone04,Nomura15}.

Recent experiments have revealed the existence of a highly anomalous metallic state near the Mott transition \cite{Ihara11,Klupp12,Potocnik14,Zadik15}.
Once the electrons are localized in the Mott phase, the electron-phonon coupling leads to a deformation of the fullerene molecule, which has been detected by IR spectroscopy in the kHz frequency range. 
On the other hand, in the conventional metallic regime, the molecules exhibit a nearly spherical shape.
In the unconventional metallic regime close to the Mott phase, called the Jahn-Teller metal (JTM) \cite{Zadik15}, a deformation of the fullerene molecules has been detected. Furthermore, recent experiments have also been performed under a magnetic field \cite{Kasahara17} and a very large upper critical field reaching 90~T has been identified in the superconducting region 
at temperatures within the JTM regime.

\begin{figure}[t]
\begin{center}
\includegraphics[width=75mm]{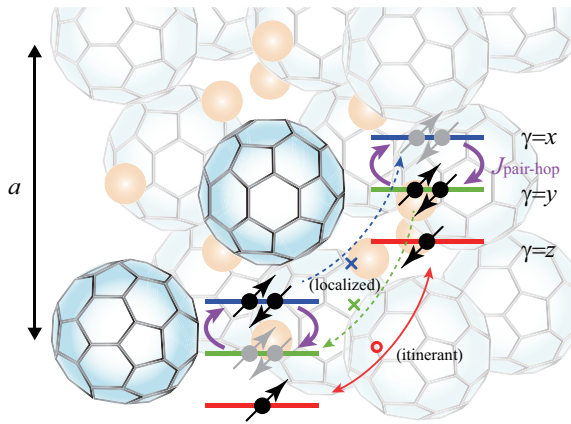}
\caption{
Schematic illustration of the SOSM state in the fcc fullerides $A_3$C$_{60}$.
Three $t_{1u}$ orbitals are schematically drawn
for two 
fullerene molecules.
The orange colored atoms represent the alkali metal $A$.
A part of this figure is reproduced from Ref.~\onlinecite{Takabayashi17}.
}
\label{fig:schematic}
\end{center}
\end{figure}

In order to clarify the microscopic origin of the JTM, the three-orbital Hubbard model has been investigated using dynamical mean-field theory \cite{Hoshino17}. This study proposed that the Jahn-Teller metal state may be interpreted as a spontaneous orbital-selective Mott (SOSM) state, in which two of the three $t_{1u}$ orbitals are spontaneously selected to become Mott insulating, while the third one stays metallic, explaining the basic properties of the JTM \cite{Hoshino17}.
Figure~\ref{fig:schematic} schematically illustrates the SOSM state for fcc fullerides.
In the figure, intra-orbital electron pairs are formed in the $x$ and $y$ orbitals, which pair-hop (with 
a transition rate $J_{\rm pair\mathchar`-hop}/\hbar$) to the other orbitals. These pairs are spatially localized, which results in a Mott insulator.
Doubly occupied orbitals are favored by 
the antiferromagnetic Hund's coupling originating from the coupling to Jahn-Teller anisotropic phonon modes \cite{Fabrizio97,Nomura15-2}.
The remaining $z$ orbital is metallic, so that the resulting state is a SOSM state. 
We note that conventional orbital-selective Mott states are realized in systems with an {\it originally} broken orbital symmetry, while here it occurs {\it spontaneously} in a system with three degenerate orbitals.
Of course, the $x$ or $y$ orbitals could equally well be selected as the metallic orbital, since the three orbitals are degenerate in a cubic structure.
Utilizing this degeneracy, it has furthermore been proposed that the ordered state can be switched on the electronic timescale by 
electric field pulses 
\cite{Werner17}. 
The orbital ordering has also been discussed in a two-dimensional system \cite{Misawa17}
and in the context of non-equilibrium superconductivity \cite{Kim16,Mazza17}.

The first task in the discussion of spontaneous symmetry breaking is to identify order parameters which are zero outside the ordered state.
Several order parameters have been proposed to characterize the SOSM state \cite{Hoshino17}, such as the orbital-dependent kinetic energy or the orbital-dependent double occupancy.
We have also characterized the ordered state by an odd-time dependent orbital moment.
However, 
the reason why these order parameters coexist is still not clear.
Furthermore, in the previous studies we considered simple particle-hole symmetric conduction bands and we do not know how this artificial choice affects the values of the order parameters.
Therefore, a systematic understanding of the order parameters of the SOSM state is still lacking.

Here, based on the realistic band structure for fulleride superconductors, we show that the recently proposed SOSM state is characterized by a nonlocal order parameter and can be regarded as an unconventional orbital ordering.
The physics behind this ordering is {\it Mottness} \cite{Phillips06}, which leads to a localized character of the electrons and an integer filling per site. 
As a consequence, as far as local quantities are concerned, the one-body order parameters vanish and the orbital-dependent two-body quantity known as the double occupancy becomes an order parameter in the SOSM state \cite{Hoshino17}.
Thus the presence of Mottness distinguishes the JTM case from the previously proposed nonlocal order parameters \cite{Pomeranchuk58, Schulz89, Fujimoto11, Hoshino11, Fu15}.

In the present study, we 
focus on spatiotemporally nonlocal single-particle quantities, which are usually more easily measurable than many-body quantities through, {\it e.g.}, transport measurements, 
while also being easily computable. 
Building on 
the previous dynamical mean-field results \cite{Hoshino17}, we characterize the SOSM state based on nonlocal order parameters, and also investigate the characteristic transport properties using the Boltzmann equation and through the optical conductivity.
Our analysis thus extends the concepts and theoretical arguments related to unconventional orders to a more realistic and practical level that can be directly connected to real materials.

This paper is organized as follows.
In the next section, we first review the nonlocal order parameters previously proposed, to clarify the novel aspect of this work.  We introduce the model that we use in Sec. III.
In Sec. IV, we analyze the order parameters, excitation spectrum, and static/dynamical transport coefficients.
We discuss in Sec. V several aspects of the unconventional diagonal orders in fullerides, and summarize the results in Sec. VI.

\section{Overview on nonlocal order parameters}

\begin{table*}
\caption{
Previously proposed electronic orderings with nonlocal order parameters in correlated systems.
All the listed order parameters are zero above the transition temperature, i.e. they are associated with a spontaneous symmetry breaking.
The components with $\bm r=\bm 0$ and $t=0$ correspond to a local quantity in space and time, respectively.
}
\label{tab:history}
\vspace{2mm}
\begin{tabular}{c|c||c}
\hline
Diagonal orders&
Examples of order parameters&
Offdiagonal-order analog
\\
\hline
\parbox[c][0.8mm][c]{0cm}{}
Pomeranchuk instability \cite{Pomeranchuk58, Yamase00, Halboth00}
/ bond-order wave \cite{Schulz89}&
$M(\bm k, \bm Q, t=0) \propto k_\mu$, $k_\mu k_\nu$, $\cos k_{\mu}$ &
anisotropic pairing\\
\hline
\parbox[c][0.8mm][c]{0cm}{}
spin nematic \cite{Fujimoto11, Fu15}& 
$M(\bm k, \bm Q, t=0) \propto k_\mu k_\nu \sg^\lambda, k_\mu \sg^\nu $ &
spin-triplet pairing\\
\hline
\parbox[c][0.8mm][c]{0cm}{}
diagonal odd-frequency order \cite{Hoshino11, Hoshino17}& 
$M(\bm r=\bm 0, \bm Q, \omega) \propto \bm \sg \omega$& 
odd-frequency pairing\\
\hline
\end{tabular}
\end{table*}

Order parameters, which characterize spontaneous symmetry breaking, can generally 
be expressed in terms of the Green functions
\begin{align}
M_{\al\al'}(\bm R_i,\bm R_j,t-t') = \delta \la \mathcal T c^\dg_{\al}(\bm R_i, t) c_{\al'}(\bm R_j,t') \ra
, \label{eq:M}
\end{align}
where $\bm R_i,\bm R_j$ are the spatial coordinates of lattice sites and $\al,\al'$ are flavor (spin/orbital) indices.
The symbol $\delta$ on the right-hand side represents the deviation from the scalar part which exist without symmetry breaking, and 
$\mathcal T$ 
is the time ordering operator.
The time dependence enters through the Heisenberg picture as $c(t) = \epn^{\imu \mathscr H t} c \epn^{-\imu \mathscr H t}$ and we note that Eq.~\eqref{eq:M} is a function of relative time only if we focus on the equilibrium state.
The flavor dependence of $M_{\al\al'}$ characterizes the symmetry breaking of the internal degrees of freedom. 
In the case of spins, the simplest examples are the Pauli matrices $M \propto\bm \sg$.

For conventional diagonal orders such as magnetic, charge and orbital orderings, the primary order parameters are considered to be local both in space and time. 
The corresponding local quantity can be written as $M_{\al\al'}(\bm R_i,\bm R_i,0)$, and its Fourier transform is given by
\begin{align}
M_{\al\al'}(\bm Q) = \frac{1}{N} \sum_{i} M_{\al\al'}(\bm R_i,\bm R_i,0) \, \epn^{-\imu \bm Q \cdot \bm R_i},
\end{align}
where $N$ is the total number of sites.
Here only the spatial modulation of this local quantity and its flavor structure matter, and the time-dependence does not necessarily enter (locality in time).
On the other hand, there also exists the possibility that the order parameter is nonlocal in space or time.
We refer to this case as {\it unconventional diagonal order}, following the terminology for superconductivity.

With the non-locality kept, the single-particle correlation functions \eqref{eq:M} can be transformed using Fourier components as 
\begin{align}
&M_{\al\al'}(\bm k,\bm Q, \omega)
= \frac{1}{N}\sum_{ij} \int \diff t 
\ 
M_{\al\al'}(\bm R_i, \bm R_j ,t)
\nonumber \\
&\hspace{20mm} \times
\, \epn^{-\imu \bm k \cdot (\bm R_i - \bm R_j)}
\, \epn^{-\imu \bm Q \cdot (\bm R_i + \bm R_j)/2}
\, \epn^{\imu \omega t},
\label{eq:M_transf}
\end{align}
where $\bm k$ and $\bm Q$ are wave vectors originating from the Fourier transform with respect to relative and center-of-mass
spatial coordinates, 
respectively.

The historically first example of a nonlocal order parameter is the Pomeranchuk instability, which results in a spontaneous deformation of the Fermi surface \cite{Pomeranchuk58, Yamase00, Halboth00}.
This mechanism is based on a weak-coupling picture with well-defined Fermi surfaces, and the order parameter describing the symmetry lowering is $\bm k$-dependent and therefore nonlocal.
A tight-binding analog of this effect has also been proposed and is called the bond-order wave \cite{Schulz89}.
If the spin-symmetry breaking is considered at the same time, and a $d$-wave like $\bm k$-space structure is assumed, the resulting state is called ``spin nematic'' \cite{Fujimoto11}. For example, $M(\bm k,\bm Q,t=0) \propto \bm d(\bm k)\cdot \bm \sg$ with $\bm d(\bm k) \propto (k_yk_z,0, 0)$ for small wave vectors and $\bm Q\neq 0$ a center-of-mass momentum to break the translational symmetry.
A $p$-wave structure in $M$ has also been proposed recently \cite{Fu15}, with the specific form $M(\bm k,0,t=0) \propto k_\mu \sg^\nu$ and $\bm Q=\bm 0$. 
These order parameters can be classified based on the point-group symmetry \cite{Hayami18}. 
In terms of nonlocality of the order parameter, these concepts are clearly related to unconventional superconductivity with $p$- and $d$-wave pairings.
Pair density waves have also been discussed \cite{Yang89,Podolsky03,Agterberg08}, but their properties are associated with the local center-of-mass coordinates and are not directly related to the nonlocality of the order parameters.

Concerning time-dependent order parameters, 
Balatsky and Abrahams considered time-dependent spin correlations in quantum spin systems, and proposed such a quantity as an order parameter for the chiral spin nematic state \cite{Balatsky95}.
Although this order is realized in spin systems and is beyond the scope of this paper,
which focuses on electronic orderings characterized by Eq.~\eqref{eq:M_transf}, its electronic analogue has also been proposed in a two-channel Kondo lattice with spontaneous channel symmetry breaking.
The order parameter is $M(\bm r=\bm 0,\bm Q=\bm 0,\omega)$ together with the spatially nonlocal one $M(\ep_{\bm k}, \bm Q=\bm 0,t=0)$ where $\ep_{\bm k}$ is the energy dispersion of the bare conduction electrons \cite{Hoshino11}.
This concept is closely related to the odd-frequency pairings with odd-time-dependent pair amplitudes \cite{Berezinskii74, Kirkpatrick91, Balatsky92, Emery92}. A brief summary of the different nonlocal order parameters is provided in Tab.~\ref{tab:history}.

In the following, we discuss in detail the nonlocal order parameter in a model for alkali-doped fullerides, where the nonlocal orbital moments inevitably
coexist 
with vanishing local orbital moments due to Mottness.
We also demonstrate that two or more nonlocal order parameters can be induced in general.

\section{Model}
Let us consider the orbital symmetry breaking in fulleride superconductors.
In the following, we focus on the zero center-of-mass momentum case ($\bm Q=\bm 0$) for ordered states.
The non-interacting Hamiltonian is given by
\begin{align}
\mathscr H_0 &= \sum_{ij}\sum_{\gm\gm'}\sum_{\sg}
t_{\gm\gm'} (\bm R_i-\bm R_j) c^\dg_{\gm\sg}(\bm R_i) c_{\gm'\sg}(\bm R_j)
\\
&= \sum_{\bm k} \sum_{\gm\gm'}\sum_{\sg}
\ep_{\gm\gm'}(\bm k) c^\dg_{\gm\sg}(\bm k) c_{\gm'\sg} (\bm k).
\end{align}
According to Ref.~\onlinecite{Nomura12}, the hopping matrix is given by
\begin{align}
t[\bm r=(\tfrac a 2 \tfrac a 2 0)]
&= 
\begin{pmatrix}
F_1&F_2&\\
F_2&F_3&\\
&&F_4
\end{pmatrix},
\\
t[\bm r=(a00)]
&= 
\begin{pmatrix}
F_5&&\\
&F_6&\\
&&F_7
\end{pmatrix},
\end{align}
where $\bm r = \bm R_i - \bm R_j$ is a relative space coordinate and $a$ is the lattice constant (see Fig.~\ref{fig:schematic}).
The other matrices are constructed by symmetry considerations.
The parameters for Rb$_3$C$_{60}$, which are considered here for simplicity, are 
$F_1=-1.6$, 
$F_2=-30.6$, 
$F_3=39.2$, 
$F_4=-15.9$, 
$F_5=-7.5$, 
$F_6=-0.8$, and
$F_7=1.5$ in units of meV \cite{Nomura12}.

Now we consider the SOSM state, where one of the three orbitals is metallic and the other two are Mott insulating with localized electrons.
To qualitatively discuss the SOSM phase, 
we introduce the self-energy for the electron Green function and an orbital dependent chemical potential and write 
\begin{align}
G^{-1}_{\gm\gm'}(\bm k,\Omega) &= \left[\Omega+\mu_\gm - \Sigma_{\gm}(\Omega)\right]\delta_{\gm\gm'} - \ep_{\gm\gm'}(\bm k),
\label{eq:G}
\end{align}
where $\Omega$ is a complex frequency and we set $\hbar = 1$.
The physical quantities on the real frequency axis $\omega$ are derived by the analytic continuation $\Omega \to \omega + \imu \eta$ with $\eta = +0$.
The local self-energy is justified in the limit of high dimensions as in the dynamical mean-field theory \cite{Georges96}. 
At sufficiently low energies
the self-energies are given by 
\begin{align}
\Sigma_{x,y}(\Omega) &= \frac{\tilde U^2}{\Omega},
\\
\Sigma_{z}(\Omega) &= 0, 
\end{align}
where $\tilde U$ is a constant with the dimension of energy that is related to the orbital-selective Mott gap.
The form of the self-energy is derived from the results obtained by the dynamical mean-field theory in the three-orbital Hubbard model with an antiferromagnetic Hund's coupling \cite{Hoshino17}, and
this parametrization suffices for the following discussion.  
In the case without particle-hole symmetry, the self energy can have the form $\Sigma_{x,y}(\Omega) = \tilde U^2/(\Omega - \Delta)$. However, $\Delta $, which is smaller than the Mott gap, can be absorbed into a shift of the chemical potential.
The quasi-particle renormalization is neglected since it only modifies the bandwidth.

Let us introduce a convenient expression for the above Green function.
By doubling the matrix size, the Green function can be written as
\begin{align}
\hat G(\bm k,\Omega) &= \left[
\begin{pmatrix}
\Omega \hat 1 + \hat \mu - \hat \ep(\bm k)& \hat U \\
\hat U & \Omega \hat 1
\end{pmatrix}
^{-1}
\right]_{11},
\end{align}
where `$11$' means that the upper-left $3\times 3$ matrix block is extracted from the generalized $6\times 6$ matrix.
We have defined 
$\hat \mu = {\rm diag\,} (\mu_x, \mu_y, \mu_z)$ and
$\hat U = {\rm diag\,} (U_x, U_y, U_z)$.
This form can be interpreted as a ``hybridization'' of electrons with the composite particles induced by strong Coulomb interactions \cite{Mancini04}.
Through diagonalization, the Green function can be rewritten as
\begin{align}
\hat G(\bm k, \Omega) &=
\check V(\bm k)_1 [\Omega \check 1 - \check \lambda(\bm k)]^{-1} [\check V^{-1}(\bm k)]_1
, \label{eq:unitary_diag}
\end{align}
where the check symbol ($\check\ $) represents the $6\times 6$ matrix.
$\check V(\bm k)_1$ is the upper half block of $\check V(\bm k)$ and is not a square matrix.
Note that the matrix sizes of $\check V(\bm k)_1$ and $[\check V^{-1}(\bm k)]_1$ are $3\times 6$ and $6\times 3$, respectively.
Thus the energy dispersion for the ``quasiparticles" in Mott insulators are described by the eigenenergies $\lambda_\al(\bm k)$.

The orbital-dependent potential $\mu_\gm$ is determined in such a way that
\begin{align}
\sum_{\sg}\la n_{\gm\sg} (\bm R_i) \ra=1
\label{eq:Mottness}
\end{align}
is satisfied for every site $i$ and orbital $\gm$.
Here we have introduced the local number operator $n_{\gm\sg}(\bm R_i) = c^\dg_{\gm\sg}(\bm R_i) c_{\gm\sg}(\bm R_i)$.
These constraints are due to the fact that a Mott insulator can be realized only for integer filling, so that Eq.~(\ref{eq:Mottness}) reflects the property of {\it Mottness}.
Since the total number per site is also fixed to three, each orbital including the metallic one must have an average filling of one electron in the orbital selective Mott state considered here.

Let us add a comment on the constraint for a related orbital symmetry broken state.
Recently, the existence of another interesting spontaneously orbital-selective state with two metallic unpaired orbitals has been revealed at higher temperatures \cite{Ishigaki18}. 
Here only one orbital is in a paired state and this pair can hop from site to site with the hopping energy $\tilde t^2/\Delta E$,  where $\tilde t$ is a renormalized hopping and $\Delta E$ is the energy needed to break the pair.
We call this state the spontaneous orbital-selective itinerant doublon (SOSID) state, to emphasize its physical nature, which is different from the SOSM state.
In this case the Mottness constraint is not 
active 
since all the electrons are delocalized, and a local order parameter (conventional orbital moment) can generally appear.
Thus, from the perspective of nonlocality of the order parameter, the SOSID state is qualitatively different from the SOSM state realized at low temperatures.
We finally note that, if the system has particle-hole symmetry, the local order parameter is zero even for the SOSID state.

\section{Analysis}

\subsection{Order parameters}
Since the orbital symmetry is clearly broken by the orbital-dependent field $\Sigma_\gm (\Omega)$, we consider the corresponding orbital moment
\begin{align}
M(\bm r) &= \frac{1}{N}\sum_{ij\gm\gm'\sg}
\lambda^8_{\gm\gm'} \la c^\dg_{\gm\sg}(\bm R_i) c_{\gm'\sg}(\bm R_j) \ra
\delta(\bm r - \bm R_i + \bm R_j),
\\
M(\bm k) &= \int \diff \bm r M(\bm r)
\epn^{-\imu \bm k\cdot \bm r},
\\
\lambda^8 &= \sqrt{\frac{1}{3}}
\begin{pmatrix}
1&&\\
&1&\\
&&-2
\end{pmatrix},
\end{align}
where the Gell-Mann matrix $\lambda^8$ describes the symmetry breaking in orbital space \cite{note1}.
With these quantities, the local orbital moment is obtained by taking the wave vector summation as
$M(\bm r=\bm 0) = N^{-1} \sum_{\bm k} M(\bm k)$. 
Its value is, however, zero due to the Mottness constraint represented by Eq.~\eqref{eq:Mottness}.
Thus, 
we need to consider the nonlocal order parameter, which distinguishes the present system from the previously discussed conventional ones.
The simplest nonlocal quantity that describes the orbital symmetry breaking is $M(\bm k=\bm 0)$.

The $\bm k=\bm 0$ and $\bm r=\bm 0$ components have the same symmetry with respect to $\bm k$-space rotations, and hence these two quantities can be simultaneously non-zero, as far as symmetry is concerned.
However, due to the Mottness constraint, the spatially local ($\bm r=\bm 0$) component becomes zero in the SOSM case, while the $\bm k=\bm 0$ component can be nonzero.
This Mottness constraint is distinct from the symmetries 
which are usually considered to impose constraints on physical quantities.

We can also consider the following nonlocal order parameter in imaginary time or the frequency domain:
\begin{align}
M(t) &= \sum_{i\gm\gm'\sg}
\lambda^8_{\gm\gm'} \la \mathcal T c^\dg_{\gm\sg}(\bm R_i) c_{\gm'\sg}(\bm R_i,t) \ra,
\\
M(\Omega=\imu\omega_n) &= \int_0^\beta \diff \tau \, M(-\imu \tau)
\, \epn^{\imu \omega_n \tau},
\label{eq:M_omega}
\end{align}
which is now chosen as spatially local.
$\beta = 1/k_{\rm B}T$ is the inverse temperature and $\omega_n = (2n+1)\pi / \beta$ is the fermionic Matsubara frequency.
The real-frequency representation can be obtained by analytic continuation $\Omega \to \omega + \imu \eta$.
In a similar manner to the discussion above, the $t=0$ component (local in time) vanishes due to Mottness, while the frequency dependent component $M(\Omega)$ can be finite.
The simplest order parameter is the $\Omega =0$ component.
In this case, the even-frequency orbital moment is mixed with the dominant odd-frequency moment, which is discussed in more detail in Sec.~IV\,B.

The short-time behavior can be characterized in another way.
As pointed out in Ref.~\onlinecite{Hoshino17}, the short-time $t$ behavior is dominated by the composite order parameters.
Namely, the $t$-linear coefficient is given by $\la [c_{i\gm\sg}^\dg,\mathscr H] c_{i\gm\sg'} \ra $, which includes two-body quantities originating from the commutation relation between the fermion operator and the interaction term.
Thus time- or frequency-dependent order parameters provide an alternative view of composite order parameters.

Let us comment on the particle-hole asymmetry in the electronic band structure.
In a previous study \cite{Hoshino17}, we have used a particle-hole symmetric density of states (DOS) with a semi-circular shape to mimic the electron conduction band.
In this case, the moment $M(\bm k= \bm 0)$ at the $\Gamma$ point also vanishes, in addition to $M(\bm r=\bm 0)$ ($=0$).
This is a peculiar feature of the particle-hole symmetric DOS, which is generally absent in real materials.
The same observation applies to the frequency dependent case: the value at $\omega=0$ also vanishes in the particle-hole symmetric model. We have checked that in the more realistic model considered here, there is no constraint which enforces $M(\bm k=\bm 0)$ and $M(\omega=0)$ (see Sec.~IV\,B).
Let us add a further comment on the $\bm k$-dependent order parameters:
the finite orbital moment at the $\Gamma$ point discussed above may be regarded as a monopole in terms of the multipolar expansion in $\bm k$-space around the $\Gamma$ point \cite{Hayami18}.
This however vanishes for the particle-hole symmetric case,
and in such a situation higher-order multipoles such as quadrupoles in $\bm k$-space must be considered to characterize the ordered state.

Another subtle issue is that there exist two or more order parameters at the same time.
Thus the question arises: which of these is the primary order parameter?
The comparison of the magnitude is difficult since there is an ambiguity in the normalization process.
Instead, we have to look at the structure of the potential that induces the spontaneous symmetry breaking, namely the anomalous self energy.
In the present case, the self-energy is local and frequency dependent (as in dynamical mean-field theory), and the primary order parameter should be regarded as the odd-frequency diagonal order parameter given in Eq.~\eqref{eq:M_omega}.
We note that this argument may change depending on the structure of the self-energy: if the anomalous self-energy is $\bm k$-dependent, the primary order parameter would be a spatially nonlocal order parameter.

\subsection{Single-particle excitation spectra}

\begin{figure*}[t]
\begin{center}
\includegraphics[width=160mm]{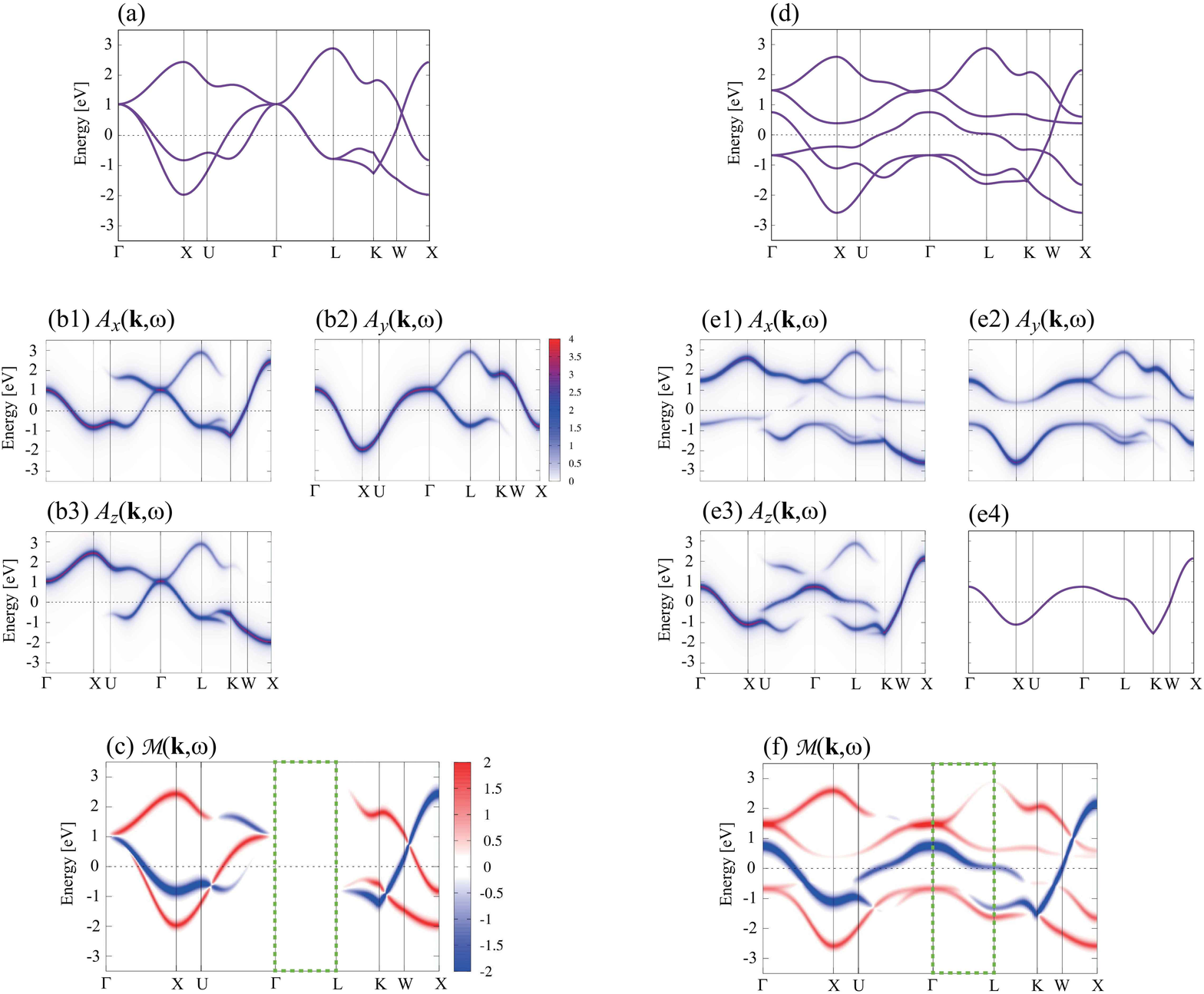}
\caption{
(a--c) Single particle spectra for the fulleride Rb$_3$C$_{60}$ without orbital ordering.
(a) bare band dispersion and (b1-b3) orbital-resolved spectra $A_{\gm=x,y,z}(\bm k,\omega)$.
The spectrally decomposed orbital moment $\mathscr M(\bm k,\omega)$ is plotted in (c).
(d,e,f) are plots similar to (a,b,c), but for the SOSM state.
The parameter $\tilde U$ is chosen as $\tilde U=0.1$ eV and the broadening factor as $\eta=8$meV.
The path in momentum space connects the points 
$\Gamma(0,0,0)$ $\to$
$X(0,1,0)$ $\to$
$U(\tfrac 1 4,1,\tfrac 1 4)$ $\to$
$\Gamma(1,1,1)$;
$\Gamma(0,0,0)$ $\to$
$L(\tfrac 1 2,\tfrac 1 2,\tfrac 1 2)$ $\to$
$K(\tfrac 3 4,\tfrac 3 4,0)$ $\to$
$W(1,\tfrac 1 2,0)$ $\to$
$X(1,0,0)$
where the coordinates for the symmetric points in $\bm k$-space are shown with the unit $\pi/a$.
}
\label{fig:spectrum}
\end{center}
\end{figure*}

Here we consider some dynamical quantities relevant for fulleride superconductors.
We define the single-particle spectrum as
\begin{align}
A_{\gm} (\bm k,\omega) &= - \frac{1}{\pi} \imag G_{\gm\gm}(\bm k,\omega+\imu \eta),
\\
A_{\gm} (\omega) &= \la A_{\gm} (\bm k,\omega) \ra_{\bm k},
\end{align}
where $\la \cdots \ra_{\bm k}$ denotes the average over $\bm k$ space, which results in a spatially local quantity. 
We also define the quantity
\begin{align}
\mathscr M (\bm k,\omega) &= \sum_{\gm} \lambda^8_{\gm\gm} A_\gm(\bm k,\omega),
\end{align}
which gives us information on the spectral decomposition
of the orbital moment.

We first show in Fig.~\ref{fig:spectrum}(a) the band structure of the fulleride superconductor Rb$_3$C$_{60}$ without any interaction effects, which 
has been obtained by density functional theory calculations
\cite{Nomura12}.
Panels (b1--b3) plot the corresponding orbital-resolved spectra $A_{x,y,z}(\bm k,\omega)$.
Since the spectral weight is not unique, it is represented as a color map with an appropriate broadening factor 
($\eta=8$ meV is taken to visualize the spectra). 
The orbital moment spectrum $\mathscr M(\bm k,\omega)$ is shown in Fig.~\ref{fig:spectrum}(c).
Although the orbital ordered moment is zero for the disordered phase considered here, $\mathscr M(\bm k,\omega)$ can be nonzero for low-symmetry $\bm k$ points other than e.g. the $\Gamma$ and L points.
In the following, we will compare these noninteracting results to those in the SOSM state.

Figure~\ref{fig:spectrum}(d) shows the 
electronic dispersion relations in the SOSM state.
The number of bands increases due to the splitting (for two orbitals) into upper and lower Hubbard bands.
Since it is not easy to understand the nature of this state from the total spectral function, we also plot orbital-resolved spectra $A_{\gm=x,y,z}(\bm k, \omega)$ in Figs.~\ref{fig:spectrum}(e1--e3).
The orbitals $x$ and $y$ exhibit a Mott gap, while the orbital $z$ is metallic.
Its Fermi surface is basically consistent with that of a system with only $z$ orbitals, whose noninteracting dispersion is plotted in Fig.~\ref{fig:spectrum}(e4).
The orbital moment spectrum is shown in Fig.~\ref{fig:spectrum}(f).
One sees that the insulating and metallic orbitals give contributions of opposite sign.
The finite values at $\Gamma$ and along the cut from $\Gamma$ to L, which is enclosed with a green dotted line in Figs.~\ref{fig:spectrum}(c) and (f), clearly illustrate the orbital symmetry breaking.

\begin{figure}[t]
\begin{center}
\includegraphics[width=80mm]{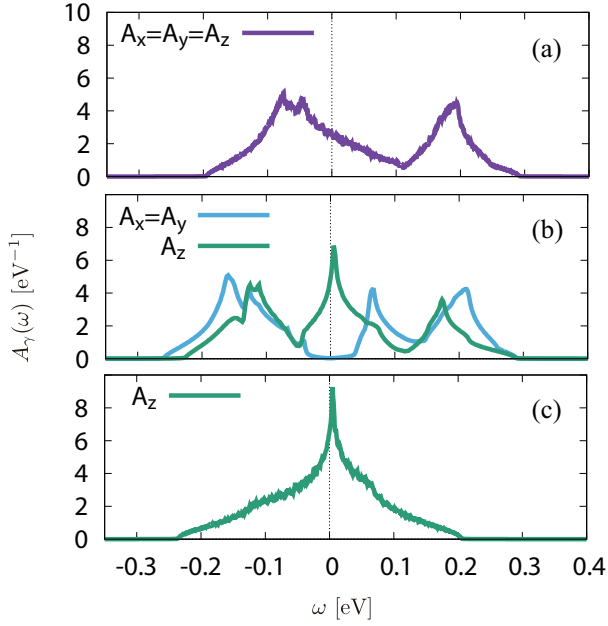}
\caption{
Wave vector integrated orbital-dependent spectra for (a) the normal metal and (b) the SOSM state.
The spectrum with only the $z$ orbital considered is shown in panel (c).
}
\label{fig:dos}
\end{center}
\end{figure}

Let us now discuss the spatially local but frequency dependent quantities.
The DOS for each orbital is plotted in Fig.~\ref{fig:dos}, where we again compare the normal metal and SOSM states. 
It follows from the comparison of panels (a) and (b) that the spectral weight is strongly reduced at low energy for the Mott insulating orbitals ($x,y$), while it increases for the $z$ orbital.
In Fig.~\ref{fig:dos}(c), we push the Hubbard bands away by taking $\tilde U\rightarrow \infty$. This leaves only the metallic $z$ orbital, whose local spectral function qualitatively reproduces the low-energy result in Fig.~\ref{fig:dos}(b).
A noteworthy feature is that the particle-hole asymmetry seen in the normal metal is strongly modified in the SOSM state, which features an almost particle-hole symmetric DOS for the $z$ orbital near the Fermi level.
Furthermore, a van-Hove singularity-like structure, as obtained e.g. in a square lattice, emerges near the Fermi level.
We will return to this point 
in Sec.~IV\,C.

Even though the value of the DOS at the Fermi level given by $\sum_{\gm}A_\gm(\omega=0)$ in the SOSM state remains comparable to that of the normal state, the system is effectively reduced to a single-orbital model. 
In this situation the multiorbital nature is lost, which is unfavorable for negative-Hund's coupling 
superconductivity, where the multiorbital nature is essential for intra-orbital pairing.
This is consistent with the reduction of the superconducting transition temperature in the JTM region \cite{Zadik15}.

\begin{figure}[t]
\begin{center}
\includegraphics[width=80mm]{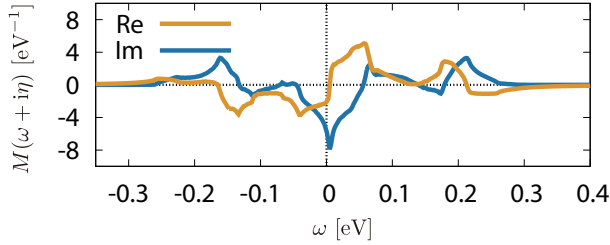}
\caption{
Spatially-local and frequency-dependent orbital ordered moment on the real axis.
}
\label{fig:M}
\end{center}
\end{figure}

We next discuss the spatially local but frequency dependent orbital moment defined in Eq.~\eqref{eq:M_omega}.
The result on the real frequency axis is obtained by the analytic continuation $M(\Omega=
\omega + \imu\eta)$. Figure~\ref{fig:M}(a) shows the real and imaginary parts of $M(\omega+\imu\eta)$.
The integration of $\imag M(\omega+\imu\eta)$ multiplied with the Fermi distribution function 
is proportional to the local order parameter. 
The data shown in the figure are consistent with a vanishing local order parameter. 
We note that, for the real part, the integral does not become zero, but this cannot be represented by a time-local quantity.

It is notable that the real part of $M(\omega+\imu\eta)$ is a nearly odd function with respect to $\omega$.
In contrast, for conventional orders, $\real M(\omega+\imu\eta)$ has an even-function shape, since the symmetry-breaking anomalous self-energy has a dominant constant contribution in the frequency domain. 
The result in Fig.~\ref{fig:M} thus provides support for the interpretation of the SOSM state as an ``odd-frequency orbital order.''
Whereas we focus on the real frequency axis in the main text, the odd-frequency nature can also be seen in the imaginary frequency domain, which is discussed in Appendix~\ref{appendix}.

\subsection{Anisotropic transport properties}

\subsubsection{Boltzmann trasport}

We now consider the transport properties of the SOSM state.
We employ the Boltzmann theory \cite{Ziman_book}, and calculate the electronic current $\bm j$ and heat current $\bm j_Q$,  which can be expressed as 
\begin{align}
\bm j &= \sg \bm E,
\\
\bm j_Q &= \kappa (-\bm \nabla T)/T.
\end{align}
The transport coefficient tensors  $\sg$ (electrical conductivity) and $\kappa$ (thermal conductivity) are given
by 
\begin{align}
&\sg 
= - e^2\tau \sum_{\al\sg} \left\la \bm v_{\bm k\al} \bm v_{\bm k\al}
\frac{\partial f(\ep_{\bm k\al})}{\partial \ep_{\bm k\al}} \right\ra_{\bm k},
\\
&\kappa
= - \frac{\tau}{T} \sum_{\al\sg} \left\la \bm v_{\bm k\al} \bm v_{\bm k\al} (\ep_{\bm k\al}-\mu)^2
\frac{\partial f(\ep_{\bm k\al})}{\partial \ep_{\bm k\al}} \right\ra_{\bm k},
\end{align}
where $f(\ep)=1/(\epn^{\ep/k_{\rm B}T}+1)$ is the Fermi distribution function.
The velocity is defined as $\bm v_{\bm k\al} = \partial \ep_{\bm k\al} / \partial \bm k$, and the eigenenergies $\ep_{\bm k\al}$ with $\al=1,2,3$ are obtained by diagonalizing $\ep_{\gm\gm'}(\bm k)$.
A single relaxation time $\tau$ is assumed. 
For the SOSM state, the Fermi surface is approximately given by that of the $z$-orbital only model.
Hence we choose $\ep_{\bm k\al} \rightarrow \ep_{zz}(\bm k)$ with $\al$-summation dropped.

We have calculated the above transport coefficients in the normal metal and SOSM state.
In the normal state they evaluate to 
\begin{align}
\frac{\sg}{\sg_0} &=
\begin{pmatrix}
11.8&&\\
&11.8&\\
&&11.8
\end{pmatrix}
,\ \ \ 
\frac{\kappa}{\kappa_0}
=
\begin{pmatrix}
0.383&&\\
&0.383&\\
&&0.383
\end{pmatrix},
\end{align}
while in the SOSM state one obtains
\begin{align}
\frac{\sg}{\sg_0} &=
\begin{pmatrix}
0.7&&\\
&8.2&\\
&&7.9
\end{pmatrix}
,\ \ \ 
\frac{\kappa}{\kappa_0}
=
\begin{pmatrix}
0.023&&\\
&0.261&\\
&&0.257
\end{pmatrix}.
\end{align}
The units are given by
$\sg_0 = \frac{e^2 \tau}{\hbar^2a}\times 1{\rm meV}$ and 
$\kappa_0 = \frac{\tau}{T\hbar ^2 a} \times (1{\rm meV})^3$ where we have explicitly written $\hbar$.
It is notable that in the SOSM case, the $y$ and $z$ components are much larger than the $x$ component.
This implies the emergence of two-dimensional transport through the spontaneous orbital symmetry breaking in fullerides.

The above observation, emergent two-dimensionality, is consistent with the DOS plotted in Fig~\ref{fig:dos}.
Namely, the DOS has a similar shape as that of a square lattice with a van-Hove singularity near the Fermi level.
This can be qualitatively understood by the large tight-binding parameter $F_3$ for the $z$ orbital, which forms a nearly square lattice within the $yz$ plane.
The slight shift of the van Hove singularity from zero seen in Fig.~\ref{fig:dos}(b,c) is due to the effect of small next-nearest-neighbor hoppings.

We note that the $t_{1u}$ molecular orbitals are sometimes interpreted as analogs of atomic $p$ orbitals, which in the case of the orbital-selective metal might incorrectly suggest a one-dimensional character, since the atomic $p$ orbitals are directed along a certain direction. 
The two-dimensionality of the SOSM state in the fullerene-based materials, demonstrated above, is a characteristic behavior originating from the complex shape of the molecular orbitals.

\subsubsection{Optical conductivity}

The characteristic features of the frequency-dependent orbital-symmetry breaking field can be seen in the optical conductivity.
We introduce the current density operator
\begin{align}
\bm j &= \frac{e}{V}\sum_{\bm k\gm\gm'\sg} \bm v_{\gm\gm'}(\bm k)  c^\dg_{\bm k\gm\sg} c_{\bm k\gm'\sg},
\end{align}
where the velocity is defined by $\bm v_{\gm\gm'}(\bm k) = \partial \ep_{\gm\gm'}(\bm k) / \partial \bm k$ and $V$ is the volume.
The dynamical current correlation function with imaginary time/frequency is given by
\begin{align}
K_{\mu \nu}(\imu \nu_m) &= V\int_0^\beta \diff \tau \la j_\mu (\tau) j_\nu \ra \epn^{\imu\nu_m \tau}
\end{align}
where $\nu_m = 2\pi m/\beta$ is a bosonic Matsubara frequency.
We assume that vertex corrections can be neglected. 
The analytic continuation $\imu \nu_m \to \omega + \imu \eta$ is analytically performed, and the final expression reads
\begin{widetext}
\begin{align}
&K_{\mu \nu} (\omega) = \frac{2 e^2}{V} \sum_{\bm k}\int \frac{\diff \omega'}{2\pi \imu} \Big(
f(\omega')
\trace [\hat v^\mu(\bm k) \hat G^{\rm A}(\bm k,\omega'-\omega) \hat v^\nu(\bm k) \hat G^{\rm A}(\bm k,\omega')]
- f(\omega')
\trace [\hat v^\mu(\bm k) \hat G^{\rm R}(\bm k,\omega') \hat v^\nu(\bm k) \hat G^{\rm R}(\bm k,\omega' + \omega)]
\nonumber \\
&\hspace{40mm}
- [f(\omega'+\omega) - f(\omega')]
\trace [\hat v^\mu(\bm k) \hat G^{\rm A}(\bm k,\omega') \hat v^\nu(\bm k) \hat G^{\rm R}(\bm k,\omega' + \omega)]
\Big),
\end{align}
\end{widetext}
where $\hat G^{\rm A}(\bm k,\omega) = \hat G(\bm k, \omega-\imu\eta)$ and $\hat G^{\rm R}(\bm k, \omega) = \hat G(\bm k, \omega+\imu\eta)$ are the advanced and retarded Green functions on the real frequency axis, respectively.
In practical calculations, it is convenient to use the diagonalized form given in Eq.~\eqref{eq:unitary_diag}.
The complex conductivity is then calculated as
\begin{align}
\sg_{\mu\nu}(\omega) &= \frac{K_{\mu \nu}(\omega)}{\imu \omega}
.
\end{align}
Thus the frequency-dependent transport properties can be studied through $\sg_{\mu\nu}(\omega)$.

\begin{figure*}[t]
\begin{center}
\includegraphics[width=130mm]{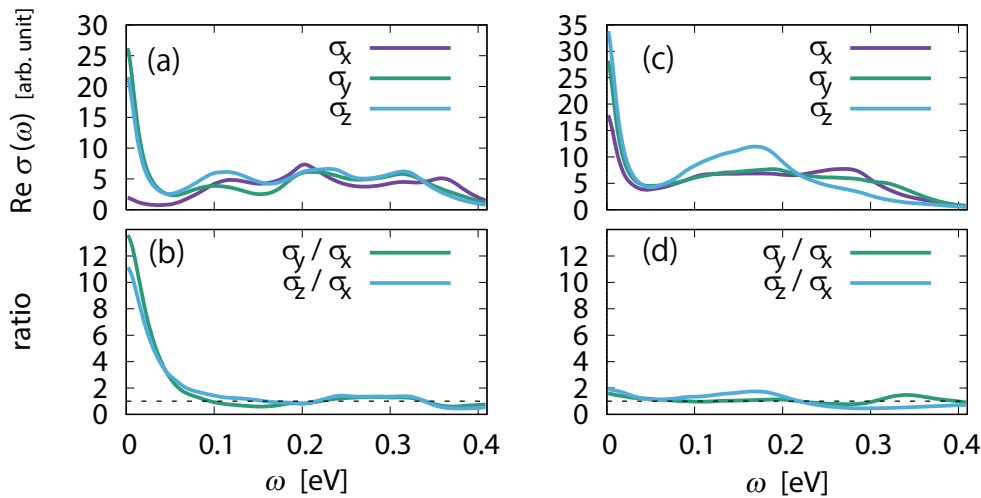}
\caption{
Frequency dependence of the optical conductivities $\sg_{xx}$, $\sg_{yy}$ and $\sg_{zz}$ for (a,b) the SOSM state 
[$\Sigma_{x,y}(\Omega) = \tilde U^2/\Omega$, $\Sigma_{z}(\Omega) = 0$] and (c,d) the system with static mean-fields 
[$\Sigma_{x,y}(\Omega) = \tilde U$, $\Sigma_{z}(\Omega) = 0$].
The damping parameter and effective Coulomb interaction are chosen as $\Gamma =15.6$ meV and $\tilde U = 0.1$ eV.
In (b,d), the dotted lines show the value corresponding to the isotropic limit realized in the original cubic phase.
}
\label{fig:opt}
\end{center}
\end{figure*}

The numerical results for the real part of the optical conductivity $\real \sg (\omega)$ are shown in Fig.~\ref{fig:opt}.
We have introduced the single-particle impurity scattering rate $\Gamma$, which is included by the replacement $\Omega \rightarrow \Omega + \imu \Gamma {\rm sgn\,} (\imag\Omega)$ in Eq.~\eqref{eq:G}, and have taken the zero-temperature limit.
The anisotropic conductivity discussed above can be seen in Fig.~\ref{fig:opt}(a):
in the low-frequency regime, the anisotropy is substantial, while it is not pronounced at high frequencies.
This behavior is more clearly visualized by looking at the ratio of conductivities shown in Fig.~\ref{fig:opt}(b).
We note that the two dimensionality appears only in the low-frequency regime.
This behavior results from the frequency-dependent orbital-symmetry breaking field.

For comparison, we also show results obtained with a frequency-{\it independent} self-energy.
Here, we simply replace $\Sigma_\gm(\Omega) = U_\gm^2/\Omega \to U_\gm$, corresponding to a static orbital-dependent mean-field (we have chosen $U_{x,y}=\tilde U \neq 0$ and $U_z = 0$).
The corresponding optical conductivity is shown in Fig.~\ref{fig:opt}(c), and the ratio characterizing the anisotropy in (d).
In strong contrast with the SOSM state the anisotropy is not enhanced at low frequencies.
Hence the frequency dependence of the symmetry-breaking fields is clearly reflected in the optical conductivity.

\section{Discussion}

We first discuss possible implications for the interpretation of experimental results on fulleride compounds. 
The emergent two-dimensionality should be a characteristic property of the JTM state, which is interpreted here as the SOSM state. 
Indeed, recent experiments show an enhancement of $H_{c2}$ in the superconducting state below the JTM regime \cite{Kasahara17}. 
If the external magnetic field is applied along the emergent two-dimensional plane, the orbital motion by the Lorentz force is not effective,
which results in higher $H_{c2}$ values. 
The use of polycrystals in the experiments implies that the above situation inevitably occurs, so that our finding accounts for the enhancement of $H_{c2}$ in the JTM region.

Even single crystals should exhibit a multidomain nature due to entropic effects, so that 
an enhancement of $H_{c2}$ can still be expected.
To align the domains in the SOSM state, it is necessary to apply an uniaxial stress in the single crystal or to produce a small enough sample with length scale below the domain size.
In this case, a characteristic magnetic field angle dependence should be observed reflecting the emergent two-dimensionality.

We also note that in real materials, the appearance of the JTM is observed as a crossover, while the theory predicts a transition with spontaneous symmetry breaking.
One possibility is that the order parameter of the SOSM state is an unconventional one and the entropy release might be small, resulting in a small anomaly at the transition point.
Another possibility is that
with multi-domains in polycrystals, an uni-axial pressure is effectively applied and turns the transition into a crossover.
We should also consider the possibility of a short-range ordered state, in which the SOSM characteristics can only be observed over a certain finite time or length scale.
In principle, this can be theoretically investigated by including spatial correlations.
Since numerical simulations based on dynamical mean-field theory and its extensions are limited to small-size clusters, spatial correlations are difficult to incorporate. 
Therefore, a Ginzburg-Landau (GL) type analysis would be a better choice.
A dynamically extended GL analysis would also be interesting for an effective description of the nonequilibrium dynamics, as discussed in Ref.~\onlinecite{Werner17}. 
A study of the physics in the presence of disorder and inhomogeneity would provide important information for a more direct connection with experiments.

\section{Summary}

We have discussed unconventional order parameters which are nonlocal in space or time and are relevant for the description of the Jahn-Teller metal in fulleride superconductors. 
The Jahn-Teller metal can be interpreted as a spontaneous orbital-selective Mott state, which is inevitably characterized by a nonlocal order parameter. In particular, we have emphasized the role of Mottness in constraining the nature of this ordered state.

We have further explored the single-particle spectra and transport properties, which show the emergence of two-dimensionality in the orbital symmetry broken state.
This is consistent with the observation of a high upper critical field in the Jahn-Teller metal regime. 
The characteristic dynamics is also found in the optical conductivity, where we have revealed the frequency-selective anisotropic transport originating from the frequency-dependent orbital-symmetry breaking fields. 

These insights should be useful for the further
exploration of strongly correlated electron phenomena in fulleride-based superconductors, and for the construction of more general concepts for ordering phenomena in condensed matter physics.

\section*{Acknowledgement}
The authors thank K. Prassides for useful discussions.
S.H. acknowledges M. Owada for making Fig.~1.
This work was supported by KAKENHI Grants No.~JP16H04021, No.~JP18K13490 (S.H.), No.~JP16H06345, and JST ERATO (JPMJER1301) (R.A.). P.W. acknowledges support from ERC Consolidator Grant No.~724103.

\appendix
\section{
Imaginary frequency representation of the orbital moment}
\label{appendix}

\begin{figure}[t]
\begin{center}
\includegraphics[width=80mm]{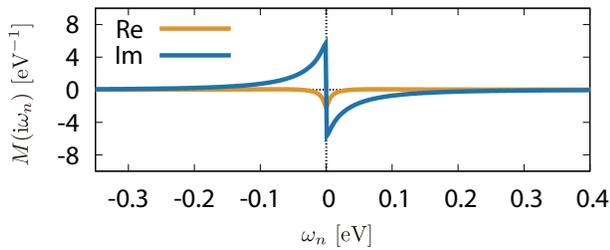}
\caption{
Spatially-local and frequency-dependent orbital ordered moment on the imaginary axis.
}
\label{fig:M2}
\end{center}
\end{figure}

The odd-frequency nature of the order parameter in the SOSM state can be even more clearly seen in the imaginary (Matsubara) frequency domain.
Figure~\ref{fig:M2} shows the real and imaginary parts of $M(\Omega = \imu\omega_n)$, which is the imaginary-frequency version of Fig.~\ref{fig:M}.
The imaginary part, which is an odd function, dominates the real part.

\begin{thebibliography}{99}
\bibitem{note1} Although another Gell-Mann matrix $\lambda^3 = {\rm diag\,}(1,-1,0)$ also describes the orbital symmetry breaking, $\lambda^8$ is shown to be energetically more stable at low temperatures \cite{Hoshino17,Ishigaki18}.
\end{thebibliography}
\end{document}